\documentclass[twoside,twocolumn,9pt]{article}
\usepackage{itujournalstyle}
\usepackage{gensymb}
\usepackage{graphicx}
\usepackage{subfigure}
\usepackage{subcaption}
\usepackage{makecell}
\usepackage{amsmath}
\usepackage{multirow}
\renewcommand{\arraystretch}{1.5}

\usepackage{caption}
\definecolor{jour_color}{rgb}{0.09, 0.24, 0.49} 
\newcommand{\Jname}{JWCC}
\newcommand{\refstyle}{ieee}

\newcommand{\TitleJournal}{REAL-TIME DIGITAL TWIN PLATFORM: A CASE STUDY ON CORE NETWORK SELECTION IN AERONAUTICAL AD-HOC NETWORKS}

\newcommand{\TurkceBaslik}{GERÇEK ZAMANLI DİJİTAL İKİZ PLATFORMU: AD-HOC HAVA AĞLARINDA ÇEKİRDEK AĞ SEÇİMİ ÜZERİNE VAKA ÇALIŞMASI }

\newcommand{\AbstractJournal}{The development of Digital Twins (DTs) is hindered by a lack of specialized, open-source solutions that can meet the demands of dynamic applications. This has caused state-of-the-art DT applications to be validated using offline data. However, this approach falls short of integrating real-time data, which is one of the most important characteristics of DTs. This can limit the validating effectiveness of DT applications in cases such as aeronautical ad-hoc networks (AANETs). Considering this, we develop a Real-Time Digital Twin Platform and implement core network selection in AANETs as a case study. In this, we implement microservice-based architecture and design a robust data pipeline. Additionally, we develop an interactive user interface using open-source tools. Using this, the platform supports real-time decision-making in the presence of data retrieval failures.} 

\newcommand{\Ozet}{Dijital İkizlerin (Dİ) geliştirilmesi için dinamik uygulamaların taleplerini karşılayabilmesine kabiliyetindeki ve açık kaynaklı çözümlerin eksiklikliklerden dolayı sekteye uğramaktadır. Bu, son teknoloji DT uygulamalarının çevrimdışı veriler kullanılarak doğrulanmasına neden olmuştur. Ancak bu yaklaşım, DT'lerin en önemli özelliklerinden biri olan gerçek zamanlı verilerin entegrasyonu konusunda yetersizdir. Bu, Ad-Hoc hava ağları (AHHA) gibi vakalarda DT uygulamalarının doğrulanmasını sınırlayabilmektedir. Bunu göz önünde bulundurarak, bu çalışmada Gerçek Zamanlı Dijital İkiz Platformu geliştirilmiş ve örnek olay olarak AANET'lerde çekirdek ağ seçimini uygulanmıştır. Bunda mikroservis tabanlı mimariye sahip dayanıklı veri işleme hattı tasarlanmıştır. Ayrıca açık kaynaklı araçlar kullanarak etkileşimli bir kullanıcı arayüzü geliştirilmiştir. Bunlar sayesinde geliştirilen platform, veri alma hataları durumunda dahi gerçek zamanlı karar almayı destekler hale getirilebilmiştir.} 
\newcommand{\Keywords}{digital twin, real-time platform, aeuronatical networks, data pipeline}
\newcommand{\AnahtarKelimeler}{dijital ikiz, gerçek zamanlı platform, hava ağları, veri işleme hattı }

\newcommand{\AuthorA}{Lal Verda Cakir}
\newcommand{\AuthorB}{Mihriban Kocak}
\newcommand{\AuthorC}{Mehmet Özdem}
\newcommand{\AuthorD}{Berk Canberk}
\newcommand{\AdresA}{School of Computing, Engineering and The Built Environment, Edinburgh Napier University, United Kingdom}
\newcommand{\AdresB}{BTS Group, Turkey}
\newcommand{\AdresC}{Department of Computer Science, Aalto University, Finland}

\newcommand{\AdresD}{Turk Telekom, Turkey}
\newcommand{\AdresE}{Department of Artificial Intelligence and Data Engineering, Istanbul Technical University, Turkey}

 \usepackage[
    backend=biber,
    style=\refstyle,
  ]{biblatex}
  \usepackage{csquotes}
\begin{document}
\justifying{
\renewcommand{\headrulewidth}{0pt}
\fancyhead[LE,RO]{\includegraphics[width=18cm, height=1.6cm]{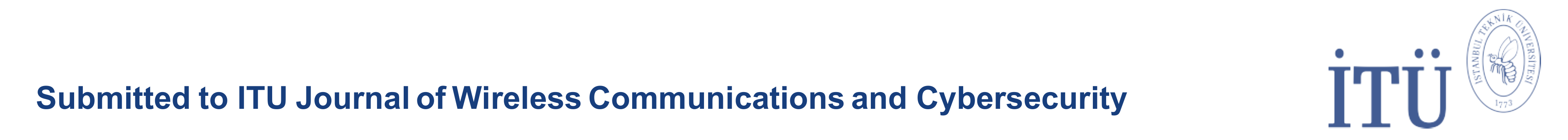}\hfill\raisebox{0pt}[0pt][0pt]}
\setlength{\headheight}{36pt}
\renewcommand{\footrule}{\hbox to \headwidth{\color{jour_color}\leaders\hrule height \footrulewidth\hfill}}
 \renewcommand{\footrulewidth}{.5pt}
\fancyfoot[LE,LO]{\textcolor{jour_color}{L. V. Cakir, M. Kocak, M. Özdem, B. Canberk}}
\fancyfoot[RE,RO]{\textcolor{jour_color}\thepage}

\sectionfont{\color{jour_color}{\sffamily\Large}}
\subsectionfont{\color{jour_color}{\normalsize}}
\subsubsectionfont{\color{jour_color}{\bf}}
\renewcommand{\figurename}{\small{Fig.}~}
\setlength{\skip\footins}{0.8cm}
\setlength{\footnotesep}{0.25cm}
\setlength{\jot}{10pt}

\twocolumn[
{\includegraphics[width=18cm]{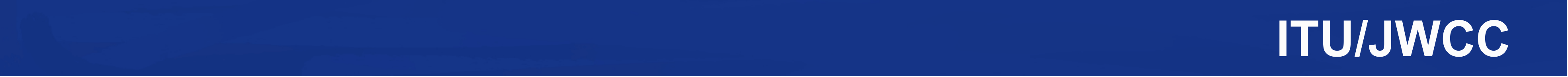}}
\bigskip
\begin{center}
\LARGE\textbf{{\color{jour_color} \TitleJournal }} \\
\bigskip

\large{\textbf{\AuthorA\textit{$^{1,2}$} \orcidA{}, \AuthorB\textit{$^{3}$} \orcidB{}, \AuthorC\textit{$^{4}$}\orcidC{}, and \AuthorD\textit{$^{1,5}$}\orcidD{}}} 

\bigskip
\large{{\textit{$^{1}$} \AdresA}}
\\
\large{{\textit{$^{2}$} \AdresB}} \\
\large{{\textit{$^{3}$} \AdresC}} \\
\large{{\textit{$^{4}$} \AdresD}} \\
\large{{\textit{$^{5}$} \AdresE}} \\
\large{{Emails: \{lal.cakir, b.canberk\}@napier.ac.uk, verda.cakir@btsgrp.com, mihriban.kocak@aalto.fi, mehmet.ozdem@turktelekom.com.tr, canberk@itu.edu.tr}}
\end{center}

\normalsize{\textbf{\color{jour_color}Abstract:} \AbstractJournal}\\
\medskip
 \normalsize{\textbf{\color{jour_color}Keywords:} \Keywords} \\

{\color{jour_color} \rule{\linewidth}{2mm}}\\
\begin{center}
\LARGE\textbf{{\color{jour_color} \TurkceBaslik}} \\ 
\end{center}
\medskip
\normalsize{\textbf{\color{jour_color}Özet: } \Ozet}\\
\medskip
\normalsize{\textbf{\color{jour_color}Anahtar Kelimeler: } \AnahtarKelimeler} \\
{\color{jour_color} \rule{\linewidth}{2mm}}\\
\medskip
\begin{center}
\large{\color{jour_color}{RESEARCH PAPER}}\\
\medskip
\large{\color{jour_color}{Corresponding Author:}} 
\large{{\AuthorA, lal.cakir@napier.ac.uk; verda.cakir@btsgrp.com}}
\medskip

{\color{jour_color} \rule{\linewidth}{.2mm}}
\bigskip
\end{center}
]

\section{Introduction}

The DT field has limited availability of platforms that cater to the diverse and complex requirements of various applications. Existing platforms are often too generic, lacking the specific features, capabilities, and programmability needed to support highly dynamic and intricate environments \cite{dtplatforms}. Additionally, many of these are not open-sourced, restricting access for researchers and developers. This shortage can significantly impact the ability to perform a comparative evaluation of DTs \cite{dt-challenge}. Due to this, existing studies on DTs have frequently relied on pre-existing data in closed environments. This lack of real-time data integration limits the ability to accurately reflect the current state and ensure the testing is valid for real-world implementations. This challenge is especially prominent in applications that require immediate responses, such as core network selection in aeronautical ad-hoc networks (AANETs). To ensure the continuous connectivity of AANETs, the incoming data must be processed, and a new decision should be made at the right time without interrupting the service. While this can be realised to some degree using the existing data within the closed environment, a new set of challenges can arise when real-time data is integrated \cite{aot, fsc_paper}. For instance, the discrepancies and irregularities in the data collection may cause the DT model to be in temporal misalignment\cite{cakir2024aienergydigitaltwining}. At this, applying estimation of these unknown factors has been a key approach \cite{spatio-temporal}. Therefore, the solutions must be tested beforehand by means of integrating real-time data to identify, address, and overcome such obstacles.

\subsection{Related Work}

The DT technology has grown in popularity due to its ability to mimic, emulate, or replicate the characteristics or specific behaviour of corresponding physical entities. This technology corresponds to creating a software model that can create real-time virtual representations of an object, process or system \cite{barricelli2019survey,t6conf}. The capabilities of DTs are supported by the rapid development of Artificial Intelligence (AI) and Machine Learning (ML) technologies \cite{dtn_survey}. With this, DTs differentiate from classic simulation models by constantly learning and adapting \cite{ddos_paper}. In this way, predictions for the future can be made and directly integrated into management's decision-making process. 

Moreover, DTs have become an important technology in the field of aviation connectivity. In this, aeronautical ad-hoc networks (AANETs) are widely used by commercial and military aviation organizations to facilitate communication. This connectivity is crucial to allow for data exchange of flight plans and weather updates between aircraft. Moreover, according to the survey results in \cite{survey} 66\% of the passengers claim that in-flight connectivity (IFC) is an essential requirement while it was seen as a luxury service in the past. Overall, AANETs are required to fulfil the high-quality internet connection demands of aviation organizations. To address this, in \cite{bilen2022proof}, the DTs are used to increase the efficiency of core network selection in WIFI-enabled in-flight connectivity (W-IFC). This study has used the recorded data from \cite{Flightradar}, which is used in sampled periods offline.  While the results acquired validated the potential of DTs in this field, the proposed architecture disregarded the streaming nature of the data. Here, developing the interaction mechanisms and related interfaces is highly important to ensure the DT model can operate accurately \cite{Zhou_Yang_Duan_Lopez_Pastor_Wu_Boucadair_Jacquenet}.

\subsection{Contributions}
Considering these, we develop the Real-Time Digital Twin Platform with a case study on core network selection in AANETs and contribute to the literature as follows:

\begin{itemize}
 
    \item We develop the real-time digital twin platform using open-source tools and leveraging microservice-based architecture. In this, we use a time-series database as the data store and develop a flask application for an interactive user interface. 
    
    \item We implement the core network selection in AANETs as a case study and use the Openskynetwork API to retrieve real-time data on the aircraft. Thanks to testing with real-time data integration, we reveal that the API can fail to return data in some instances due to unknown factors. This is a highly possible case in real-world scenarios that can occur due to factors such as congestion, failures and backlogs. 
    
    \item The robust data pipeline is developed using a stream processing engine, Kapacitor, to support real-time decision-making. In this, the feature scaling, clustering, and recommendation operations of core network selection are performed streamingly. We design this pipeline so that it can handle the case of an extract operation failing to return or empty data. For this, we implement the projection module, which is activated to calculate the information on the aircraft. This robust design can allow core network selections to continue without disruption to the service.

\end{itemize}

\section{Real-Time Digital Twin Platform}
\begin{table}[h]
   \centering
   \setlength{\tabcolsep}{12pt} 
\renewcommand{\arraystretch}{1.5} 
    \begin{tabular}{ |l|l| }
    \hline
      \textbf{Field Name} & \textbf{Description}  \\
      \hline
    $icao24$ & Transponder Address\\
    $callsign$ & Callsign of Aircraft \\
    $origin\_country$ & Country Name \\
    $time\_position$  & Timestamp of Position \\
    $last\_contact$  & Timestamp for Update\\
    $longitude$  & WGS-84 Longitude \\
    $latitude$  & WGS-84 Latitude \\
    $baro\_altitude$  & Barometric Altitude (m) \\
    $on\_ground$  &  If surface position report \\
    $velocity$  & Ground Velocity of aircraft (m/s) \\
    $true\_track$  & Direction of Aircraft\\ 
    $vertical\_rate$  & vertical velocity  of Aircraft (m/s) \\
    $geo\_altitude$  & Geometric Altitude \\
    $spi$  & Flight Status \\
    $category$  & Aircraft Category\\
    
\hline
    \end{tabular}
\caption{Data Retrieved from OpenSky Network API \cite{opensky_rest_api}}
\label{table:aircraft_data}
\end{table}

\subsection{Physical Layer}\label{sec:model}

\begin{figure*}[ht]
\centering
  \includegraphics[width=\linewidth]{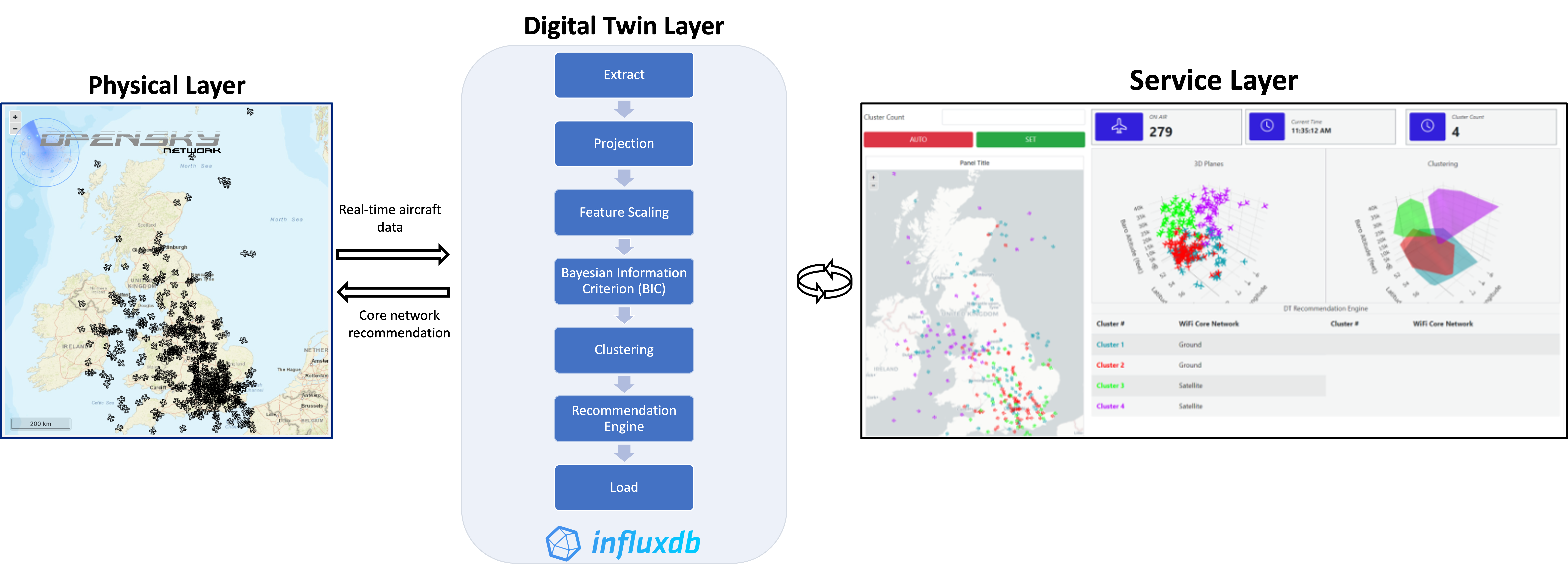}
  \caption{Real-Time Digital Twin Platform}
  \label{fgr:dt-architecture}
\end{figure*}
The Physical Layer consists of the entities and processes in which their DT is created. In the scope of this article, we consider the core network selection in the AANETs use case and specifically chose the UK area because of its active and diverse air traffic, which carries the main characteristics of AANETS. Accordingly, we retrieve real-time air traffic data from this layer using the OpenSky Network \cite{schafer2014opensky}. The details of the data is given in the Table \ref{table:aircraft_data}.

\subsection{Digital Twin (DT) Layer}
At the DT Layer, the DTs are created and maintained using the real-time data coming from the Physical Layer and their data is stored in the time-series database, InfluxDB \cite{influxdb2023}. In our DT Platform, this process is handled by the Robust Data Pipeline design, which is responsible for extracting, transforming and loading (ETL) operations and handling the disruption in the data flow. 

\subsubsection{Robust Data Pipeline}

The real-time core network selection in our DT Implementation is performed according to the following data pipeline as illustrated in the Figure~\ref{fgr:dt-architecture}.

\begin{enumerate}

    \item \textbf{Extract}: The data that the details given in Table~\ref{table:aircraft_data} is retrieved repetitively from the API. Here, in case the incoming data is empty, the last records on the aircrafts are extracted from the time-series database.
    \item \textbf{Transform}: The incoming data undergoes the following operation steps.
        \begin{enumerate}
        \item \textbf{Projection:} The timestamp of the position ($time\_position$) in the data may not be the same for each aircraft. To account for that, the location is corrected using the folowing equations:
        \begin{align}
            \label{eq:equation1}
            \Delta x &= velocity*cos(true\_track)*\Delta t \\
            \label{eq:equation2}
            \Delta y &= velocity*sin(true\_track)*\Delta t \\
            \label{eq:equation3}
            \Delta z &= vertical\_rate*\Delta t
        \end{align}
        \item \textbf{Feature scaling:} We scale features to a range of 0 to 1 using the Min-Max Scaler from the Scikit-learn library \cite{pedregosa2011scikit}.
        \item \textbf{Bayesian Information Criterion (BIC):} Before clustering we decide the optimal number of clusters (k) using the Bayesian Information Criterion (BIC) method which is claimed as a superior alternative to the elbow method \cite{Schubert_2023}.
        \item \textbf{Clustering:} We perform clustering using the K-means algorithm thanks to its simplicity and effectiveness in similarity-based clustering \cite{macqueen1967some}. We use KMeans class from the Scikit-learn library \cite{pedregosa2011scikit} over the scaled data in the previous step.
        \item \textbf{Recommendation:} We perform reccomendation for the core network selection based on the methodology introduced in the \cite{bilen2022proof}.
    \end{enumerate}
    \item \textbf{Load}: The results are loaded into the time-series database with the buckets, Physical and DT. Here, the Physical bucket stores only the data retrieved from the API while the DT bucket stores the data after projection, cluster assignment and the corresponding recommendation. Here, the data is represented as line protocols that are in the following form: 
    \begin{equation}
        measurement\ tag,tag1,...\ field,field3,...\ timestamp
    \end{equation}

\end{enumerate}

\subsection{Service Layer}
In the Service Layer of our Real-Time Digital Twin Platform, we develop a web application interface using  Flask framework \cite{flask} and Grafana \cite{grafana2019} and Plotly \cite{plotly2015}. By this, the monitoring, clustering results, and recommendations are visualised in an interactive dashboard as shown in Figure~\ref{fgr:dt-architecture}. Within this, aircraft clusters are displayed through three different dashboards in which the aircraft are colour-coded based on their respective clusters. The first dashboard portrays aircraft on a map with their 2D coordinates (longitude and latitude) determining their positions. The second dashboard presents the aircraft clusters with a 3D scatter plot. The third dashboard plots the distinctive 3D coverage areas of aircraft clusters, encompassing all aircraft belonging to each specific cluster. Moreover, the recommendations for the clusters for their core network selection are given.





\section{Performance Evaluation}
\begin{figure}[h]
    \centering
    \includegraphics[width=\linewidth]{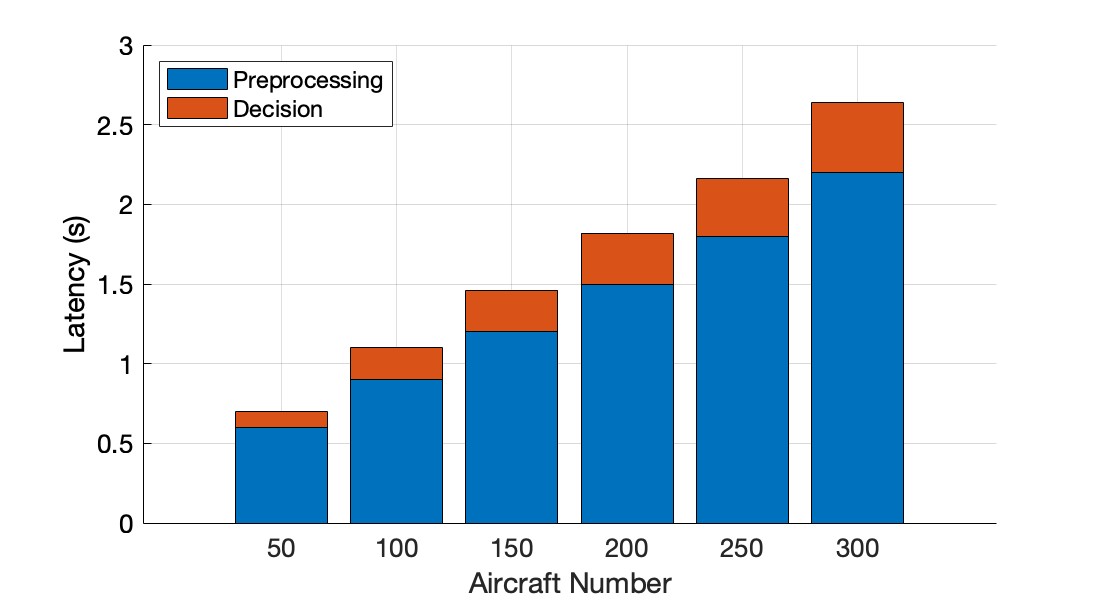}
    \caption{Evaluation of the latency for preprocessing and decision making}
    \label{fig:perf1}
\end{figure}

We evaluate the Real-Time Digital Twin Platform by using the latency, clustering accuracy and cluster change rate metrics. Here, the latency is categorized as the preprocessing latency and the decision latency, corresponding to the time taken for the data integration and the decision-making in the recommendation engine. To withstand the dynamic of the automatical environment, the platform should operate with low latency. To evaluate that, we measure the decision latency of our implementation, as shown in Figure~\ref{fig:perf1}. Here, as the number of aircraft increases, the delay also increases due to the processed data volume. Moreover, while the latency incurred from just the decision-making process shows similar results to the initial study \cite{bilen2022proof}, we observe that the preprocessing of the data mainly dominates the total latency. 

\begin{figure}[h]
    \centering
    \includegraphics[width=\linewidth]{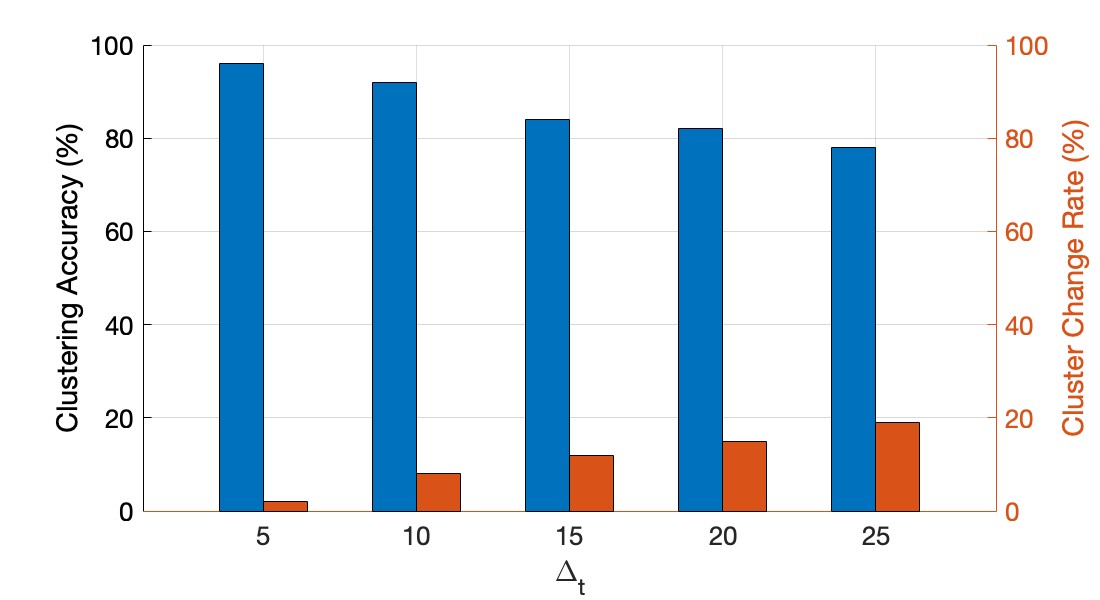}
    \caption{Evaluation of the data preprocessing pipeline for decision making}
    \label{fig:perf2}
\end{figure}
Moreover, in Figure~\ref{fig:perf2}, the accuracy of the cluster assignment when the projection is applied is shown with the corresponding cluster change rate. Here, the clustering accuracy is calculated by using the projection step for an instance of extracted data and comparing the results for the next extraction step. As for the cluster change rate, it is the percentage of cluster assignment that has to be changed compared to the baseline. In the core network selection, the cluster change occurs to provide better connectivity for the aircraft, and the projection is applied to ensure that the necessary cluster change decision can be made if the data cannot be retrieved. Here, the results show that the platform can make these decisions thanks to this robust mechanism. However, it is also observed that, as the $\Delta t$ increases, the accuracy of the results lowers, which is due to Equations~\ref{eq:equation1}-\ref{eq:equation3} not considering the aircraft's acceleration and the emergence of the new aircraft. Meanwhile, using projection in further time differences causes a higher cluster change rate. This shows that the wrong cluster assignments affect the decision-making process's effectiveness. 

\section{Conclusion}
In this article, we implement a real-time digital twin platform for core network selection in aeronautical ad-hoc networks. We design a robust data pipeline to integrate real-time aircraft data, handle the missing data, and decide on the core network selection. Moreover, we develop an interactive user interface using open-source tools. Correspondingly, we analyze the results for latency, accuracy of cluster assignment and cluster change rate. This reveals that the proposed platform can integrate real-time data and apply decision-making with delays lower than 3 seconds. Moreover, the use of projections enables the handling of missing data.

\section*{Acknowledgements}
This work is supported by The Scientific and Technological Research Council of Turkey (TUBITAK) 1515 Frontier R\&D Laboratories Support Program for BTS Advanced AI Hub: BTS Autonomous Networks and Data Innovation Lab. Project 5239903.



\printbibliography 

@inproceedings{schafer2014opensky,
    author = {Matthias Schafer and Matthias Strohmeier and Vincent Lenders and Ivan Martinovic and Matthias Wilhelm},
    title = {Bringing up OpenSky: A large-scale ADS-B sensor network for research},
    booktitle = {IPSN-14 Proceedings of the 13th International Symposium on Information Processing in Sensor Networks},
    year = {2014},
    doi = {10.1109/IPSN.2014.6846743}
}

@article{bilen2022proof,
  title={A proof of concept on digital twin-controlled WIFI Core Network selection for in-flight connectivity},
  author={Bilen, T. and Ak, E. and Bal, B. and Canberk, B.},
  journal={IEEE Communications Standards Magazine},
  volume={6},
  number={3},
  pages={60--68},
  year={2022}
}

@article{barricelli2019survey,
  title={A survey on Digital Twin: Definitions, characteristics, applications, and design implications},
  author={Barricelli, B. R. and Casiraghi, E. and Fogli, D.},
  journal={IEEE Access},
  volume={7},
  pages={167653--167671},
  year={2019}
}

@misc{opensky_rest_api,
    title = {Opensky Rest Api},
    howpublished = {\url{https://openskynetwork.github.io/opensky-api/rest.html}},
    note = {Accessed: 2023-05-19},
    year = {2023},
}

@article{pedregosa2011scikit,
    author = {F. Pedregosa and G. Varoquaux and A. Gramfort and V. Michel and B. Thirion and O. Grisel and M. Blondel and P. Prettenhofer and R. Weiss and V. Dubourg and J. Vanderplas and A. Passos and D. Cournapeau and M. Brucher and M. Perrot and E. Duchesnay},
    title = {Scikit-learn: Machine learning in Python},
    journal = {Journal of Machine Learning Research},
    volume = {12},
    pages = {2825--2830},
    year = {2011}
}

@article{Schubert_2023, 
    title={Stop using the elbow criterion for k-means and how to choose the number of clusters instead}, 
    volume={25}, 
    DOI={10.1145/3606274.3606278}, 
    number={1}, 
    journal={ACM SIGKDD Explorations Newsletter}, author={Schubert, Erich}, 
    year={2023}, 
    month={Jun}, 
    pages={36–42}}

@inproceedings{macqueen1967some,
  title={Some Methods for Classification and Analysis of Multivariate Observations},
  author={MacQueen, J. B.},
  booktitle={Proceedings of the Fifth Berkeley Symposium on Mathematical Statistics and Probability},
  volume={1},
  number={14},
  year={1967},
  pages={281--297}
}

@book{flask,
  title={Flask Web Development: Developing Web Applications with Python},
  author={Grinberg, Miguel},
  year={2018},
  publisher={O'Reilly Media, Inc.}
}

@misc{influxdb2023,
  title={Get Started with InfluxDB OSS 2.7},
  howpublished={\url{https://docs.influxdata.com/influxdb/v2.7/}},
  note={Accessed: May 21, 2023}
}

@misc{grafana2019,
  author={{Grafana Labs}},
  title={Grafana Documentation},
  year={2019},
  month={Jul 25},
  howpublished={\url{https://grafana.com/docs/}},
  note={Online; accessed July 25, 2019}
}

@misc{plotly2015,
  author={{Plotly Technologies Inc.}},
  title={Collaborative Data Science},
  year={2015},
  url={https://plot.ly}
}

@ARTICLE{dtn_survey,
  author={Wu, Yiwen and Zhang, Ke and Zhang, Yan},
  journal={IEEE Internet of Things Journal}, 
  title={Digital Twin Networks: A Survey}, 
  year={2021},
  volume={8},
  number={18},
  pages={13789-13804},
  doi={10.1109/JIOT.2021.3079510}}

@misc{Flightradar, title={Live flight tracker - real-time flight tracker map}, url={http://FlightRadar24.com}, journal={Flightradar24}, author={Flightradar}, language={en} }

@misc{survey,
    title = {Inflight Connectivity Survey – Global Whitepaper I August 2018}, url={https://www.inmarsat.com/content/dam/inmarsat/corporate/documents/aviation/insights/2018/Inmarsat%20Aviation%202018%20Inflight%20Connectivity%20Survey%20ENG.pdf},
    author={{Inmarsat Aviation}}
}

@misc{Zhou_Yang_Duan_Lopez_Pastor_Wu_Boucadair_Jacquenet, title={Network Digital Twin: Concepts and reference architecture}, url={https://datatracker.ietf.org/doc/draft-irtf-nmrg-network-digital-twin-arch/}, journal={IETF Datatracker}, author={Zhou, Cheng and Yang, Hongwei and Duan, Xiaodong and Lopez, Diego and Pastor, Antonio and Wu, Qin and Boucadair, Mohamed and Jacquenet, Christian}, language={en} }

@ARTICLE{dtplatforms,
  author={Lehner, Daniel and Pfeiffer, Jérôme and Tinsel, Erik-Felix and Strljic, Matthias Milan and Sint, Sabine and Vierhauser, Michael and Wortmann, Andreas and Wimmer, Manuel},
  journal={IEEE Software}, 
  title={Digital Twin Platforms: Requirements, Capabilities, and Future Prospects}, 
  year={2022},
  volume={39},
  number={2},
  pages={53-61},
  doi={10.1109/MS.2021.3133795}}

@ARTICLE{dt-challenge,
  author={Fuller, Aidan and Fan, Zhong and Day, Charles and Barlow, Chris},
  journal={IEEE Access}, 
  title={Digital Twin: Enabling Technologies, Challenges and Open Research}, 
  year={2020},
  volume={8},
  number={},
  pages={108952-108971},
  doi={10.1109/ACCESS.2020.2998358}}

@ARTICLE{aot,
  author={Duran, Kübra and Özdem, Mehmet and Hoang, Trang and Duong, Trung Q. and Canberk, Berk},
  journal={IEEE Internet of Things Magazine}, 
  title={Age of Twin (AoT): A New Digital Twin Qualifier for 6G Ecosystem}, 
  year={2023},
  volume={6},
  number={4},
  pages={138-143},
  doi={10.1109/IOTM.001.2300113}}

@misc{cakir2024aienergydigitaltwining,
      title={AI in Energy Digital Twining: A Reinforcement Learning-based Adaptive Digital Twin Model for Green Cities}, 
      author={Lal Verda Cakir and Kubra Duran and Craig Thomson and Matthew Broadbent and Berk Canberk},
      year={2024},
      eprint={2401.16449},
      archivePrefix={arXiv},
      primaryClass={cs.LG},
      url={https://arxiv.org/abs/2401.16449}, 
}

@INPROCEEDINGS{fsc_paper,
  author={Ak, Elif and Canberk, Berk},
  booktitle={GLOBECOM 2020 - 2020 IEEE Global Communications Conference}, 
  title={FSC: Two-Scale AI-Driven Fair Sensitivity Control for 802.11ax Networks}, 
  year={2020},
  volume={},
  number={},
  pages={1-6},
  doi={10.1109/GLOBECOM42002.2020.9322153}}

@ARTICLE{ddos_paper,
  author={Yigit, Yagmur and Bal, Bahadir and Karameseoglu, Aytac and Duong, Trung Q. and Canberk, Berk},
  journal={IEEE Communications Standards Magazine}, 
  title={Digital Twin-Enabled Intelligent DDoS Detection Mechanism for Autonomous Core Networks}, 
  year={2022},
  volume={6},
  number={3},
  pages={38-44},
  doi={10.1109/MCOMSTD.0001.2100022}}

@INPROCEEDINGS{spatio-temporal,
  author={Gutierrez-Estevez, David M. and Canberk, Berk and Akyildiz, Ian F.},
  booktitle={2012 IEEE 23rd International Symposium on Personal, Indoor and Mobile Radio Communications - (PIMRC)}, 
  title={Spatio-temporal estimation for interference management in femtocell networks}, 
  year={2012},
  volume={},
  number={},
  pages={1137-1142},
  doi={10.1109/PIMRC.2012.6362517}}

@ARTICLE{t6conf,
  author={Ak, Elif and Duran, Kübra and Dobre, Octavia A. and Duong, Trung Q. and Canberk, Berk},
  journal={IEEE Communications Magazine}, 
  title={T6CONF: Digital Twin Networking Framework for IPv6-Enabled Net-Zero Smart Cities}, 
  year={2023},
  volume={61},
  number={3},
  pages={36-42},
  doi={10.1109/MCOM.003.2200315}}

}
\end{document}